\documentstyle[12pt,psfig]{article}
\def\gsim{\stackrel{>}{{}_\sim}}
\def\lsim{\stackrel{<}{{}_\sim}}

\def\be{\begin{equation}}       
\def\ee{\end{equation}}
\def\bear{\be\begin{array}}      
\def\eear{\end{array}\ee}
\def\bea{\begin{eqnarray}}
\def\eea{\end{eqnarray}}

\def\ie{{\it i.e.}}
\def\etal{{\it et al.}}
\def\half{{\textstyle{1 \over 2}}}

\def\bold#1{\setbox0=\hbox{$#1$}%
     \kern-.025em\copy0\kern-\wd0
     \kern.05em\copy0\kern-\wd0
     \kern-.025em\raise.0433em\box0 }
\begin{document}
\catcode`@=11
\newtoks\@stequation
\def\subequations{\refstepcounter{equation}%
\edef\@savedequation{\the\c@equation}%
  \@stequation=\expandafter{\theequation}
  \edef\@savedtheequation{\the\@stequation}
  \edef\oldtheequation{\theequation}%
  \setcounter{equation}{0}%
  \def\theequation{\oldtheequation\alph{equation}}}
\def\endsubequations{\setcounter{equation}{\@savedequation}%
  \@stequation=\expandafter{\@savedtheequation}%
  \edef\theequation{\the\@stequation}\global\@ignoretrue

\noindent}
\catcode`@=12
\begin{titlepage}

\begin{flushright}
IC/96/251 \\
IFIC/97-56 \\
hep--ph--9708288 \\
August 1997
\end{flushright}

\vspace*{4mm}

\begin{center}
{\Huge Extracting SUSY Parameters from}\\
\vspace*{5mm}
{\Huge Selectron and Chargino Production${}^{\dag}$}\\[14mm]

{\large{\bf Marco Aurelio D\'\i az${}^{1,2}$}\\
\hspace{3cm}\\
{${}^{1}$\small Departamento de F\'\i sica Te\'orica,  
IFIC--CSIC, Universidad de Valencia }\\
{\small Burjassot, Valencia 46100, Spain} \\
\hspace{3cm}\\
{${}^{2}$\small High Energy Section,  
International Centre for Theoretical Physics }\\
{\small Trieste 34100, Italy} \\}
\end{center}
\vspace{4mm}
\begin{abstract}

We review the extraction of fundamental supersymmetric parameters from 
experimental observables related to the detection of charginos and 
selectrons at $e^+e^-$ colliders. We consider supergravity models with
universal scalar and gaugino masses and radiatively broken electroweak
symmetry. Two scenarios are considered: (a) The lightest chargino is
light enough to be produced at LEP2, and (b) the right handed selectron
is light enough to be produced at LEP2. We show how the validity 
of supergravity models can be tested even if experimental errors are
large. Interesting differences between the spectrum in the two scenarios 
are pointed out. 

\end{abstract}
\vskip 5.cm
\noindent ${}^{\dag}$ Talk given at the Trieste Conference on ``Quarks and 
Leptons: Masses and Mixings'', ICTP, 7-11 October 1996, Trieste, Italy.

\end{titlepage}

\setcounter{page}{1}

\section{Introduction}

One of the main problems in particle physics today is the mass generation
mechanism. The Higgs mechanism \cite{Higgs} was proposed as a way of 
generating mass to the gauge bosons and 
fermions, nevertheless, within the Standard 
Model (SM) the mass of the Higgs scalar is unstable under radiative 
corrections. Supersymmetry (SUSY) \cite{SUSY} is a symmetry which 
protects the scalar masses against the quantum corrections, although, 
this symmetry must be broken to be in agreement with the experimental 
observations. The most popular supersymmetric extension of the SM
is the Minimal Supersymmetric Standard Model (MSSM) whose particle
content includes a scalar partner of all known fermions, a fermionic 
partner of all known gauge bosons, and two Higgs doublets plus their
fermionic partners \cite{MSSMrep}. The MSSM conserves the $R$--parity,
which means susy particles are always produced in pairs at the accelerators,
and that all susy particles eventually decay into the lightest 
supersymmetric particle (generally the lightest neutralino), which is 
stable.

Presently at LEP an extensive search of supersymmetric particles is
been performed. Negative searches impose a lower bound on the mass of
the susy particles. In this way, $m_{\tilde e^{\pm}_R}>58$ GeV
if $m_{\tilde e^{\pm}_R}-m_{\chi^0_1}>3$ GeV \cite{selexp,selchaexp} 
is the latest published bound on the right selectron, and 
$m_{\chi^{\pm}_1}>75$ GeV if $m_{\chi^{\pm}_1}-m_{\chi^0_1}>10$ 
GeV \cite{selchaexp,chaexp} is the latest published bound on the lightest 
chargino.
In this talk I will concentrate in the last two particles. In particular,
I want to investigate the problem of extracting the supersymmetric
parameters out of the experimental observables associated with the 
discovery of charginos and selectrons. 

\section{Supergravity Models}

The supergravity (SUGRA) motivated version of the MSSM is particularly 
interesting because of its predictability. In general it is assumed the 
universality of scalar and gaugino masses at the unification scale 
($M_X\approx 10^{16}$ GeV), where the gauge coupling constants 
unification is achieved, and these masses evolve differently down to 
the electroweak scale. One of the Higgs masses squared is driven towards 
negative values, due to the heaviness of the top quark, and in this way, 
the electroweak symmetry is broken radiatively.
Imposing that the renormalised tadpoles are equal to zero we can find 
the one--loop corrected minimisation condition of the Higgs potential
\cite{marcotadpoles}
\begin{equation}
\left[m_{1H}^2+{\textstyle{1\over{v_1}}}\widetilde T_1^{\overline{MS}}(Q)
+\half m_Z^2c_{2\beta}\right]c_{\beta}^2=
\left[m_{2H}^2+{\textstyle{1\over{v_2}}}\widetilde T_2^{\overline{MS}}(Q)
-\half m_Z^2c_{2\beta}\right]s_{\beta}^2
\label{eq:OneLoopMin}
\end{equation}
where $\widetilde T_i^{\overline{MS}}(Q)$ are the one--loop tadpoles, and
the dependence on the arbitrary scale $Q$ has been omitted from all
the running parameters. We include in the loops contributions from 
top and bottom quarks and squarks.

The independent parameters defined at the unification scale which specify 
the model are the scalar mass $m_0$, the gaugino mass $M_{1/2}$, the 
trilinear mass $A$, the bilinear mass $B$, and the supersymmetric Higgs
mass $\mu$.
In Minimal Supergravity it is a common practice to impose the relation
$A=B+m_0$ at the unification scale, and we use this relation in the
study of selectron pair production. Nevertheless, the relation $B=2m_0$ 
at the unification scale appears in models proposed to solve the 
$\mu$--problem \cite{b2m0}, and we adopt it in the study of the 
chargino production. These kind of boundary conditions eliminates the 
bilinear soft mass $B$ from the group of independent parameters.

\begin{figure}
\centerline{\protect\hbox{\psfig{file=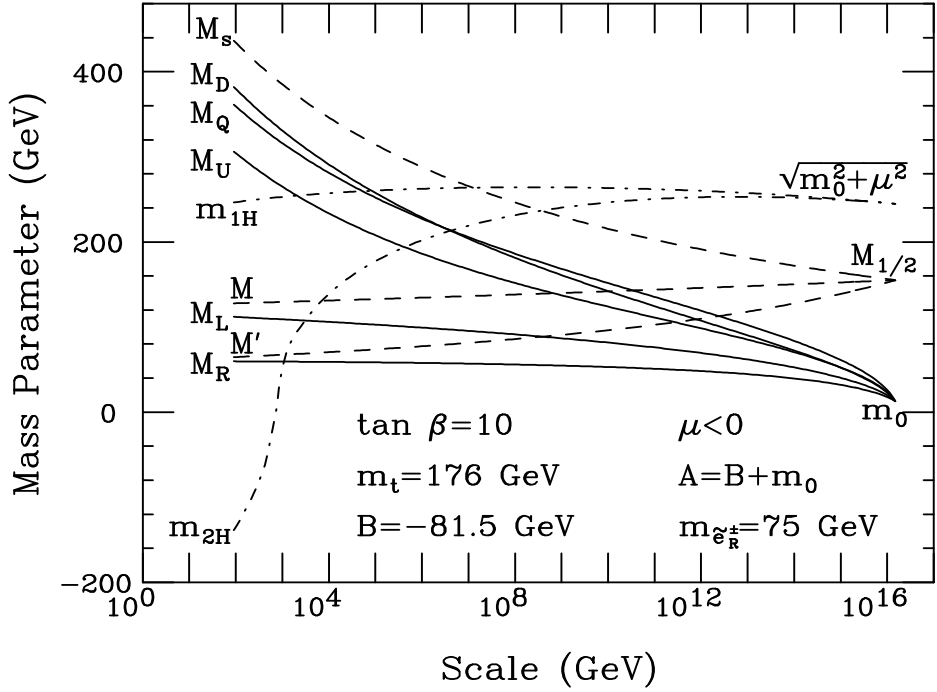,height=11cm,width=1.0\textwidth,angle=90}}}
\caption{
Scale evolution of the different soft supersymmetry breaking terms
in the supergravity lagrangian.} 
\label{fig:runpar}
\end{figure} 
In Fig.~\ref{fig:runpar} we show the running of different mass parameters 
with the arbitrary mass scale in Minimal Subtraction ($\overline{MS}$).
In solid lines we have some scalar masses, which are degenerated at the
unification scale and equal to $m_0$. Squark and slepton soft masses 
are represented by $3\times3$ mass matrices, and we plot the third
diagonal element corresponding to the third generation. We have the
left squark mass $M_Q$, the right up--type squark mass $M_U$, the 
right down--type squark mass $M_D$, the left slepton mass $M_L$, and
the right charged slepton mass $M_R$. Due to strong interactions, the
squark masses are typically larger than slepton masses. The value
of $m_0$ has been chosen to produce a selectron with 
$m_{\tilde e^{\pm}_R}=75$ GeV. In dashed lines we plot the gaugino 
masses $M_s$, $M$, and $M'$ corresponding to the groups $SU(3)$, $SU(2)$, 
and $U(1)$ respectively. They are also degenerated
at the unification scale and equal to $M_{1/2}$. Again, strong interactions
make evolve the gluino mass $M_s$ to higher values compared to the 
wino and bino masses. Thus, in supergravity, the best candidates to be 
found first at the accelerators are the sleptons, the charginos, and the
neutralinos. Finally, in dot--dashed lines we plot the Higgs masses
$m_{1H}$ and $m_{2H}$, which have the common value of $\sqrt{m_0^2+\mu^2}$
at $M_X$. In the case of the latest mass parameter, $m_{2H}^2$ is
driven to negative values as we approach to the electroweak scale. In 
that case, we plot $-\sqrt{|m_{2H}^2|}$. For this figure we adopt the 
relation $A=B+m_0$ valid at the unification scale and consider $\mu<0$.

\section{Selectron Production}

In this section we assume that the right selectron $\tilde e_R^{\pm}$
is light enough to be produced at LEP2, and study the determination
of the fundamental parameters of the supergravity model from the
experimental determination of the selectron mass, its total production
cross section, and the mass of the lightest neutralino \cite{BdeCyyo}. 
We consider the relation $A=B+m_0$ at the unification scale. We calculate
the total production of a pair of right selectrons including the
contribution from $Z$--boson and photon in the $s$--channel, and from
neutralinos in the $t$--channel.

\begin{figure}
\centerline{\protect\hbox{\psfig{file=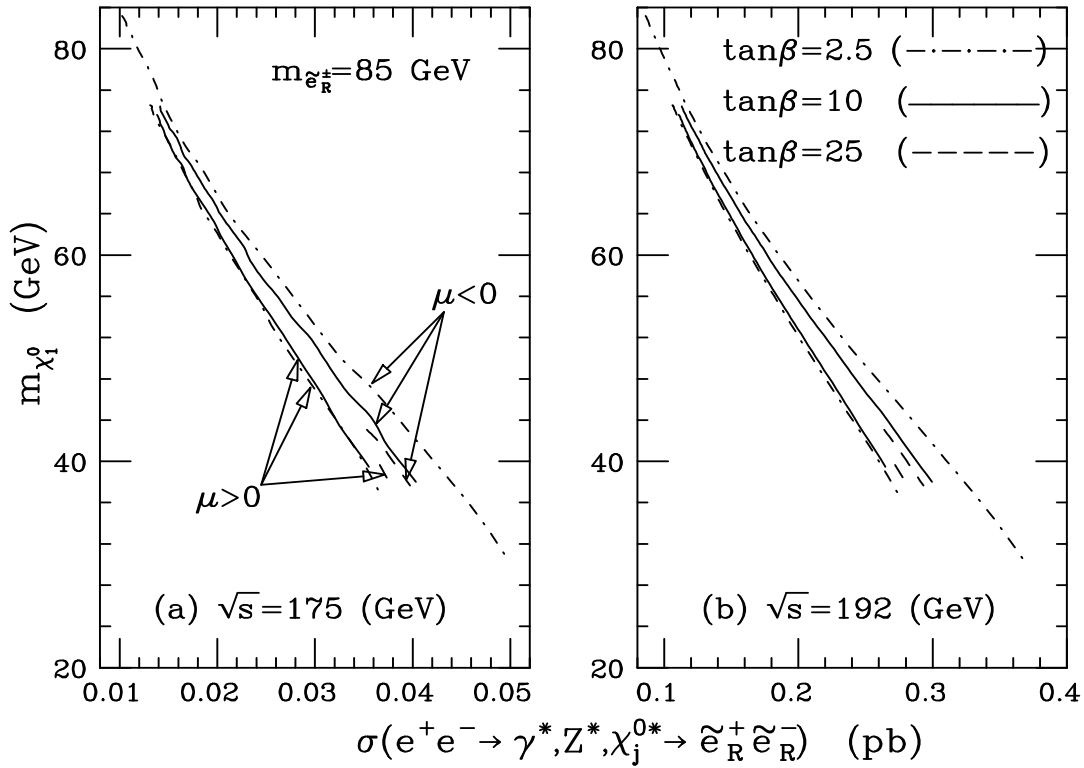,height=11cm,width=1.0\textwidth,angle=90}}}
\caption{Relation between the lightest neutralino mass and the total 
production cross section of a pair of right selectrons for different 
choices of $\tan\beta$ and the sign of $\mu$. Two values of the center
of mass energy are displayed: (a) 175 GeV and (b) 192 GeV.} 
\label{fig:s85ne}
\end{figure} 
In Fig.~\ref{fig:s85ne} we plot the relation between the lightest neutralino
mass $m_{\tilde\chi_1^0}$ and the total production cross section of a
pair of right selectrons 
$\sigma(e^+e^-\longrightarrow\tilde e_R^+\tilde e_R^-)$ for a constant
value of the selectron mass $m_{\tilde e_R^{\pm}}=85$ GeV. Curves with
$\tan\beta=2.5$, 10, 25, and both signs of $\mu$ are displayed.
The center of mass energy is (a) $\sqrt{s}=175$ GeV and (b) $\sqrt{s}=192$ 
GeV. It is obvious from the figure that the allowed points in the
$m_{\tilde\chi_1^0}-\sigma$ plane are restricted to a narrow region.
This implies that the observation of a pair of selectrons at LEP2 will
validate or ruled out the supergravity model depending on whether the
experimental results lie in the allowed region or not. It is interesting 
to notice that the neutralino mass can be close to the selectron mass 
selected here, specially for low $\tan\beta$ and $\mu<0$. This small mass 
difference decrease the efficiency of the detection of selectrons, and 
therefore, mass lower bounds become weaker in the case of non--observation. 
It can also be appreciated from the figure that it is harder to distinguish
different values of $\tan\beta$ when $\mu>0$. The way of using this 
figure is simple: once a selectron is observed, the measurement of its
mass, its production cross section, and the mass of the lightest
neutralino coming from its decay, will single out a point over one
of the curves in Fig.~\ref{fig:s85ne}. This in turn will enable the 
determination of $\tan\beta$ and the sign of $\mu$. 

\begin{figure}
\centerline{\protect\hbox{\psfig{file=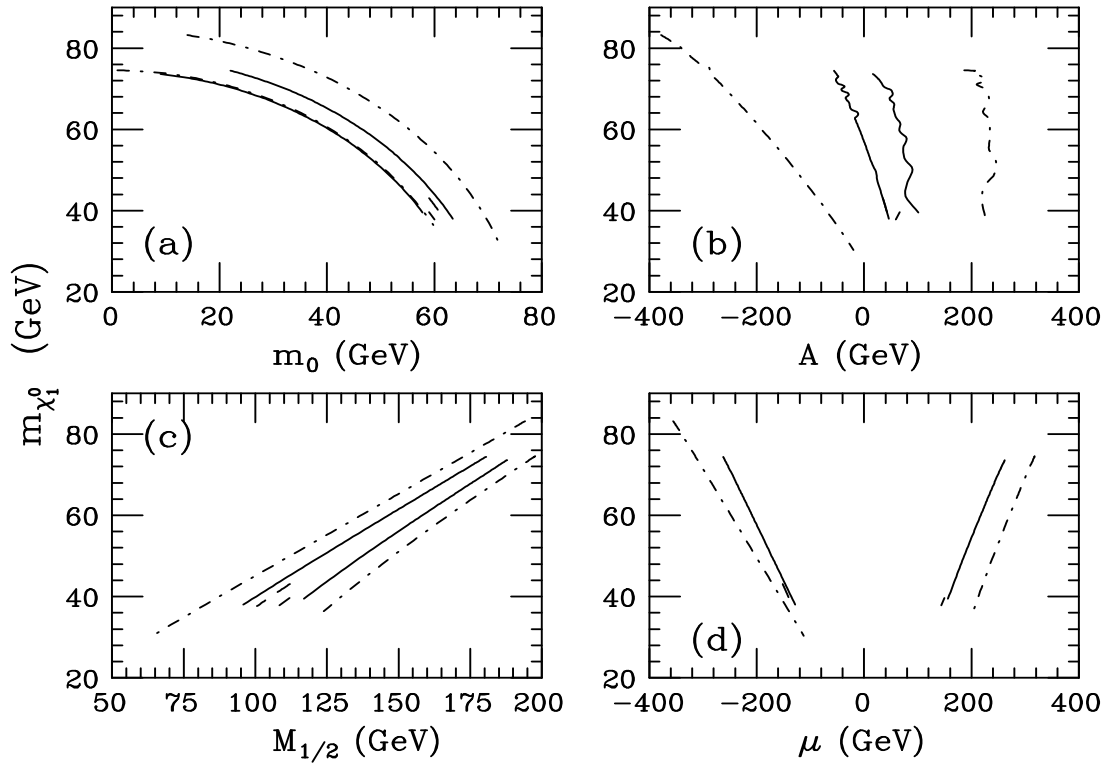,height=11cm,width=1.0\textwidth,angle=90}}}
\caption{Relation between the lightest neutralino mass 
$m_{\tilde\chi_1^0}$ and the fundamental parameters of the theory (a) 
the universal scalar mass $m_0$, (b) the universal trilinear coupling 
$A$, (c) the universal gaugino mass $M_{1/2}$, and (d) the 
supersymmetric Higgs mass $\mu$.} 
\label{fig:s85pm}
\end{figure}
For the scenario described in Fig.~\ref{fig:s85ne} we can relate to any
point in one of the curves, the value of any of the fundamental parameters 
of the theory. In Fig.~\ref{fig:s85pm} we choose to plot the relation 
between the lightest neutralino mass and the following parameters:
(a) the universal scalar mass $m_0$, the universal trilinear coupling $A$,
(c) the universal gaugino mass $M_{1/2}$, and (d) the supersymmetric
Higgs mass $\mu$. The advantage of choosing to plot the lightest
neutralino mass instead of the cross section is that we make 
Fig.~\ref{fig:s85pm} independent of the center of mass energy, \ie, 
Fig.~\ref{fig:s85pm} and the following two are valid for {\sl any}
center of mass energy. The scalar mass $m_0$ is rather low $0\le m_0<80$
GeV, and $m_0=0$ can be accommodated if $\tan\beta$ is small and $M_{1/2}$
is large. An inverse relation can be appreciated between $m_0$ and 
$M_{1/2}$, and this is because the selectron mass receive contributions 
from scalar masses as well as from gaugino masses. Large values of 
the gaugino mass are allowed: $60<M_{1/2}<200$ GeV. Both signs of $A$
are obtained, and it is mildly correlated with the sign of $\mu$. Low 
values of $|\mu|$ are not allowed because either the chargino mass or 
the stau mass is too low.

\begin{figure}
\centerline{\protect\hbox{\psfig{file=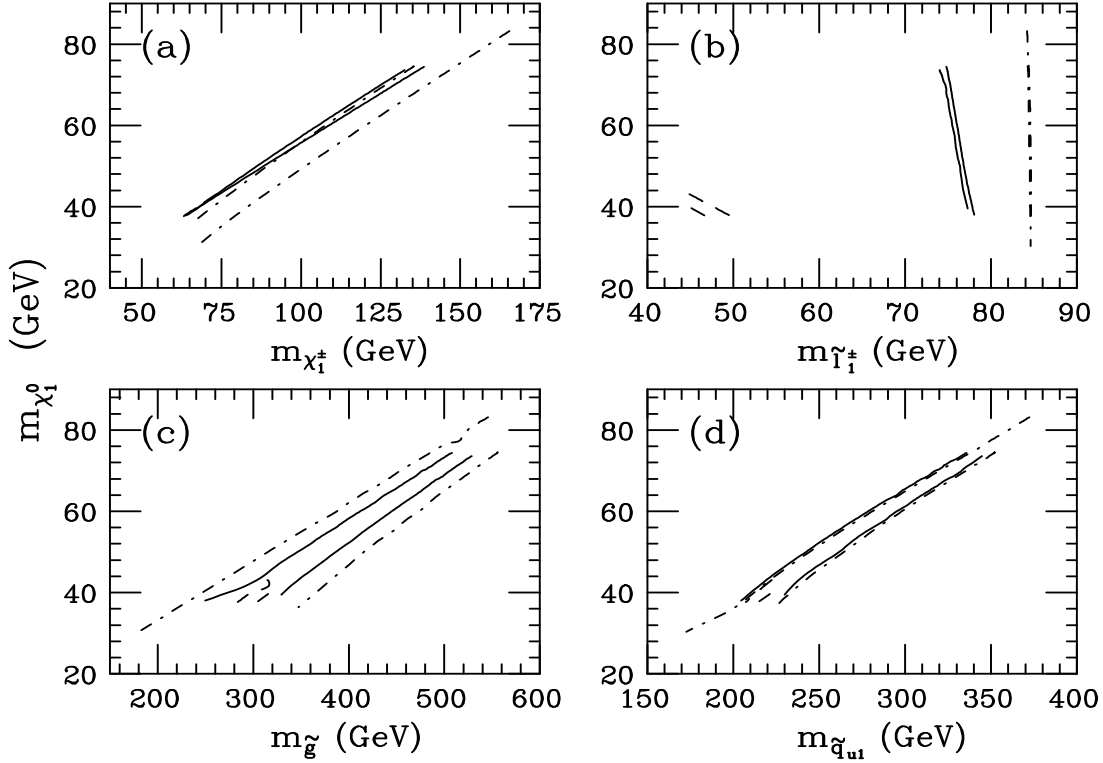,height=11cm,width=1.0\textwidth,angle=90}}}
\caption{Relation between the lightest neutralino mass 
$m_{\tilde\chi_1^0}$ and the following sparticle masses: (a) lightest 
chargino $m_{\tilde\chi_1^{\pm}}$, (b) lightest charged slepton 
$m_{\tilde l_1}$ (the stau), (c) gluino $m_{\tilde g}$, and (d) the 
lightest up--type squark $m_{\tilde q{u1}}$ (mainly stop).} 
\label{fig:s85sp}
\end{figure} 

In Fig.~\ref{fig:s85sp} we plot as a function of the lightest neutralino 
mass (a) the lightest chargino mass 
$m_{\tilde\chi_1^{\pm}}$, (b) the lightest charged slepton mass
$m_{\tilde l_1}$ (the stau), (c) the gluino mass $m_{\tilde g}$, and 
(d) the lightest up--type squark mass $m_{\tilde q{u1}}$ (mainly stop).
It is clear from the figure that $\tilde\chi_1^{\pm}$, $\tilde g$, and
$\tilde t_1$ are strongly correlated with $M_{1/2}$. The lightest
chargino can be as heavy as 170 GeV. The gluino satisfy 
$180<m_{\tilde g}<560$ GeV. We do not consider the light gluino 
scenario because it is ruled out is this class of supergravity models
\cite{lightgluino}. The top squark is bounded by $170<m_{\tilde t_1}<380$ 
GeV. It is interesting to see the stau mass in Fig.~\ref{fig:s85sp}b,
because it is strongly correlated with the value of $\tan\beta$. Since
its mass is smaller than 90 GeV, it can be pair produced at LEP2 and
a measurement of its mass can be used to determinate the value of 
$\tan\beta$.

\begin{figure}
\centerline{\protect\hbox{\psfig{file=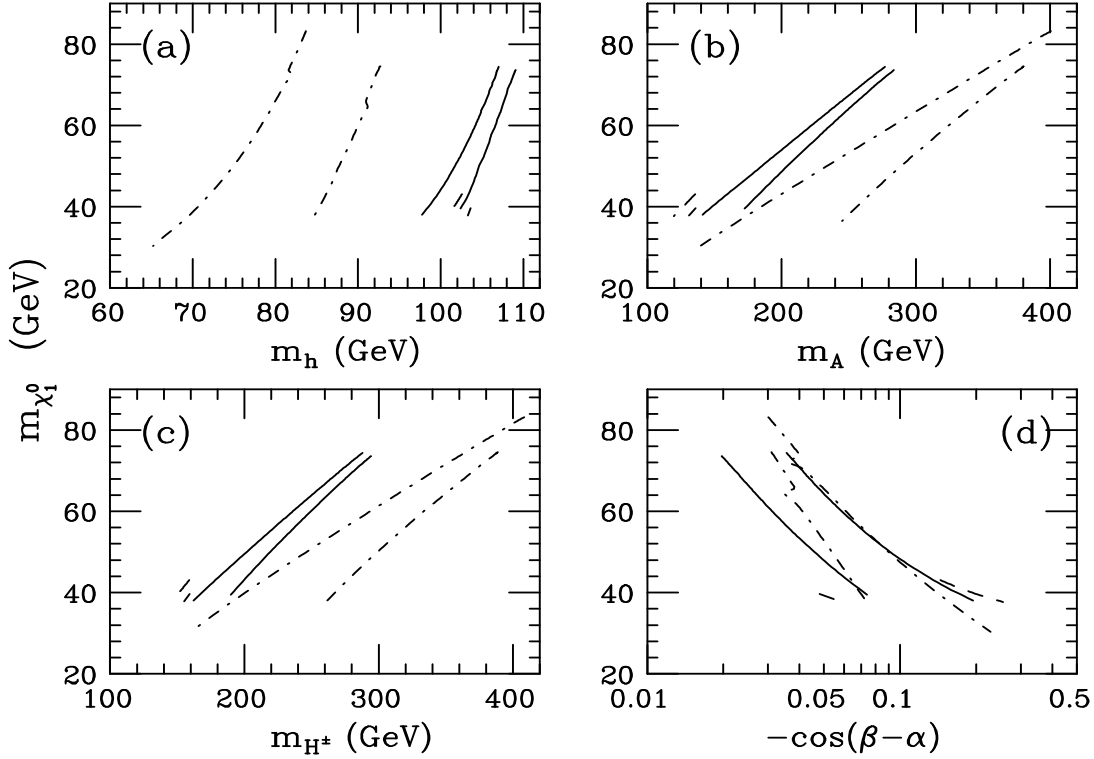,height=11cm,width=1.0\textwidth,angle=90}}}
\caption{Relation between the lightest neutralino mass 
$m_{\tilde\chi_1^0}$ and the following parameters of the Higgs sector: 
(a) lightest CP--even Higgs mass $m_h$, (b) CP--odd Higgs mass $m_A$, 
(c) charged Higgs mass $m_{H^{\pm}}$, and (d) the parameter 
$-\cos(\beta-\alpha)$.} 
\label{fig:s85hs}
\end{figure}

The Higgs sector is analysed in Fig.~\ref{fig:s85hs}. The lightest CP-even
Higgs mass $m_h$ include one--loop radiative corrections which have
been proved to be very important \cite{mhneutral}. Here it
is calculated using the method developed in \cite{alpha}.
This mass satisfy $64<m_h<110$ GeV, therefore, if it is light enough it
may be detected at LEP2, specially if the center of mass energy
$\sqrt{s}=200$ GeV is achieved. If the Higgs is detected first, the
value of its mass can be used to distinguish the supergravity model 
from the SM, since a gap emerges between the upper limit of $m_h$ in 
the first model and the lower limit of $m_{H_{SM}}$ in the second model
\cite{Vanderbilt}. In addition, it can be seen a strong dependence of 
$m_h$ on $\tan\beta$ if this parameter is small. Therefore, if $m_h$
is measured in addition to selectron detection, it can be useful to
determine the value of $\tan\beta$ in a region ($\tan\beta\gsim 2$) where
the stau mass is less sensible to this parameter. 
In Fig.~\ref{fig:s85hs}b
we have the CP-odd Higgs mass $m_A$, which is the pole mass, and it is
determined with the relation 
\begin{equation}
m_A^2={{B\mu}\over{s_{\beta}c_{\beta}}}(Q)-{{s_{\beta}^2}\over{v_1}}
\widetilde T_1^{\overline{MS}}(Q)-{{c_{\beta}^2}\over{v_2}}\widetilde
T_2^{\overline{MS}}(Q)+\widetilde A_{AA}^{\overline{MS}}(m_A^2,Q)
\label{eq:mAtadpoles}
\end{equation}
where $\widetilde A_{AA}^{\overline{MS}}(m_A^2,Q)$ is the finite self
energy of the CP--odd Higgs $A$ in the $\overline{MS}$ scheme, evaluated
at external momenta $p^2=m_A^2$. This self energy depends on the arbitrary
scale $Q$, but the overall dependence of the pole mass $m_A$ cancels
at the one--loop level. The allowed range of the CP--odd Higgs mass in 
this scenario is $120<m_A<400$ GeV. The charged Higgs mass is plotted in
Fig.~\ref{fig:s85hs}c and satisfy $150<m_{H^{\pm}}<410$ GeV. It includes 
radiative corrections \cite{mhcharged}, nevertheless, for the values
of $\tan\beta$ allowed in this scenario, quantum corrections are small.
In Fig.~\ref{fig:s85hs}d we plot the parameter $-\cos(\beta-\alpha)$.
We note that $\sin(\beta-\alpha)$ is the $ZZh$ coupling relative to the 
SM coupling $ZZH_{SM}$, therefore, $\cos(\beta-\alpha)$ close to zero 
implies the lightest CP--even Higgs $h$ has SM--like couplings. In this
scenario, $|\cos(\beta-\alpha)|<0.3$.

\section{Chargino Production}

In this section we assume that the lightest chargino $\tilde\chi_1^{\pm}$
is light enough to be produced at LEP2, and study the determination
of the fundamental parameters of the supergravity model from the
experimental determination of the chargino mass, its total production
cross section, and the mass of the lightest neutralino 
\cite{Steveyyo,chaothers}. We consider the relation $B=2m_0$ at the 
unification scale. A comparison between this choice and the minimal
supergravity relation $A=B+m_0$ is made in ref.~\cite{marcotadpoles}.
We calculate the total production of a pair of light charginos including 
the contribution from $Z$--boson and photon in the $s$--channel, and from
electron--sneutrino $\tilde\nu_e$ in the $t$--channel.

\begin{figure}
\centerline{\protect\hbox{\psfig{file=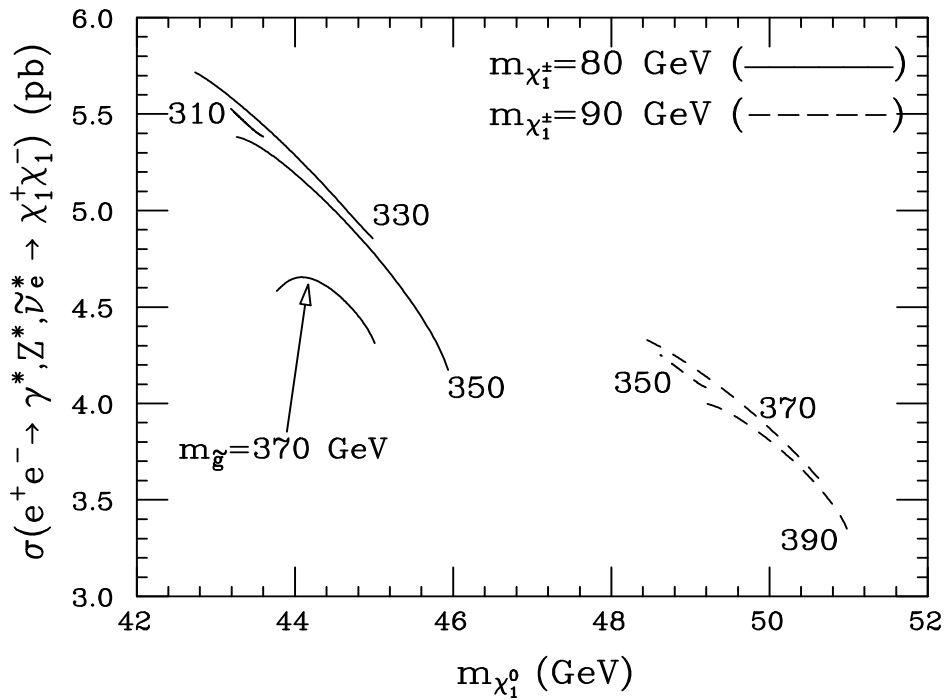,height=11cm,width=1.0\textwidth,angle=90}}}
\caption{
Total production cross section of a pair of light charginos in electron
positron annihilation as a function of the lightest neutralino mass.} 
\label{fig:c8090ne}
\end{figure} 
Total production cross section of a pair of light charginos in electron
positron annihilation as a function of the lightest neutralino mass
is plotted in Fig.~\ref{fig:c8090ne}. We take a constant value of the 
chargino mass: $m_{\chi^{\pm}_1}=80$ GeV (solid lines) and 90 GeV (dashed 
lines), in a supergravity model based in the boundary condition $B=2m_0$ 
valid at the unification scale. This implies that only one sign of $\mu$
is allowed, $\mu>0$, because $m_A^2$ must be positive. Different curves of 
the same type are labelled by the gluino mass, and the center of mass energy 
is $\sqrt{s}=192$ GeV. As in the case with selectrons, given the chargino
mass, the allowed region is small. Therefore, the observation of a 
pair of charginos at LEP2 will validate or ruled out this supergravity
model depending on whether the experimental results lie in the allowed
region or not. Contrary to the selectron case, the lightest neutralino mass
cannot be close to the chargino mass. In fact, $m_{\tilde\chi_1^0}$
is about one half of $m_{\tilde\chi_1^{\pm}}$, and this implies that in 
this class of models the the region of parameter space with low efficiency 
for the detection of charginos is avoided. The way to use 
Fig.~\ref{fig:c8090ne} is simple. A measurement of the chargino mass, its
productions cross section, and the mass of the neutralino mass, which comes
from the $\tilde\chi_1^{\pm}$ decay mode, will single out a curve in
Fig.~\ref{fig:c8090ne}, and therefore a value of $m_{\tilde g}$. Of
course, experimental errors will translate into errors in the determination 
of the gluino mass. 

\begin{figure}
\centerline{\protect\hbox{\psfig{file=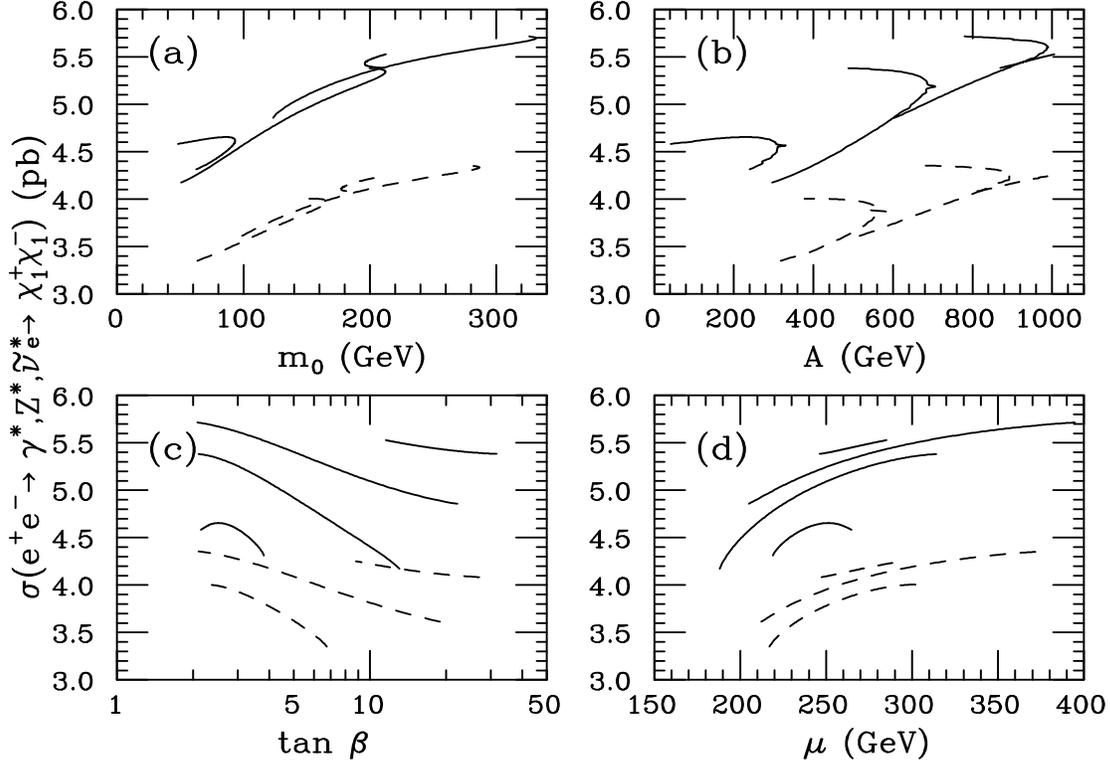,height=11cm,width=1.0\textwidth,angle=90}}}
\caption{
Relation between the total production cross section of two charginos 
$\tilde\chi_1^{\pm}$ and the fundamental parameters of the theory 
(a) the universal scalar mass $m_0$, (b) the universal trilinear
coupling  $A$, (c) $\tan\beta$, and (d) the supersymmetric Higgs mass 
$\mu$.} 
\label{fig:c8090pm}
\end{figure} 
For the scenario described in Fig.~\ref{fig:c8090ne} we plot in 
Fig.~\ref{fig:c8090pm} the relation between the total production cross 
section of two charginos and the fundamental parameters of the theory 
(a) the universal scalar mass $m_0$, (b) the universal trilinear mass 
parameter $A$, (c) the ratio between vacuum expectation values $\tan\beta$, 
and (d) the Higgs mass parameter $\mu$. In Fig.~\ref{fig:c8090pm}a we see
that the scalar mass can take large values: $50<m_0<330$ GeV while
the gaugino mass $M_{1/2}$ is kept low in order to have a light chargino.
In this sense, the light chargino scenario is complementary to the 
light selectron scenario presented in the previous section. In 
Fig.~\ref{fig:c8090pm}b we plot the parameter $A$, and appreciate that
only solutions with positive $A$ are obtained. We see that the smaller the 
gluino mass the larger the $A$ parameter, which can be as large as 1 TeV.
The parameter $\tan\beta$ is given in Fig.~\ref{fig:c8090pm}c, whose
allowed values are $2\lsim\tan\beta\lsim30$. Most of the time, the 
chargino production cross section decreases when $\tan\beta$ increases.
The last fundamental parameter we plot is the supersymmetric Higgs mass
$\mu$ in Fig.~\ref{fig:c8090pm}d. Only positive values are allowed with
$180<\mu<400$ GeV. In general, we appreciate that the total cross section
increases when $\mu$ increases. The way to use Fig.~\ref{fig:c8090pm}
and the following two figures is as follows: once the chargino and the 
gluino masses are known, a curve is singled out, and with it and the value 
of the total cross section, any parameter can be read from its corresponding 
figure.

\begin{figure}
\centerline{\protect\hbox{\psfig{file=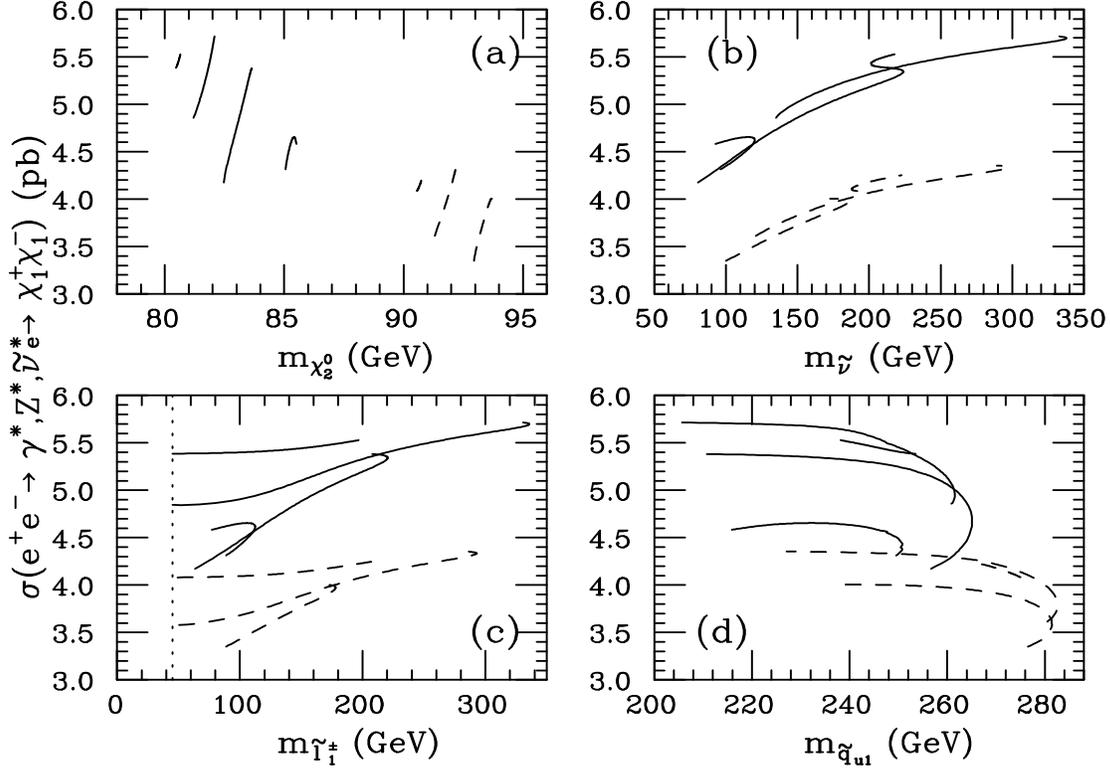,height=11cm,width=1.0\textwidth,angle=90}}}
\caption{
Relation between the total production cross section of two charginos 
$\tilde\chi_1^{\pm}$ and the following sparticle masses: (a) second 
lightest neutralino $m_{\tilde\chi_2^0}$, (b) sneutrino $m_{\tilde\nu}$ 
(the three sneutrino spices are practically degenerated), (c) lightest 
charged slepton $m_{\tilde l_1}$ (the stau), and (d) the lightest 
up--type squark $m_{\tilde q{u1}}$ (mainly stop).} 
\label{fig:c8090sp}
\end{figure} 
In Fig.~\ref{fig:c8090sp} we plot as a function of the total production
cross section of a pair of light charginos (a) the second lightest 
neutralino mass $m_{\tilde\chi_2^0}$, (b) the sneutrino mass $m_{\tilde\nu}$,
(c) the lightest charged slepton (stau) mass $m_{\tilde\tau_1^{\pm}}$, and
(d) the lightest up--type squark (mainly stop) mass $m_{\tilde t_1}$.
{}From Fig.~\ref{fig:c8090sp}a we appreciate that the second lightest
neutralino $\tilde\chi_2^0$ has a mass low enough to be produced at LEP2,
either in association with $\tilde\chi_1^0$ or pair produced. Furthermore,
in this plot the different curves labelled by the gluino mass are well
differentiated, and therefore, a measurement of $m_{\tilde\chi_2^0}$ can
help to determine $m_{\tilde g}$, which is essential for the determination
of the fundamental parameters of the model. The three sneutrino spices are 
in practice degenerated, and in Fig.~\ref{fig:c8090sp}b we plot this
common mass $m_{\tilde\nu}$. It is strongly correlated with $m_0$, and
it is clear from the figure that the electron--sneutrino contribution
to the cross section is negative, and the lighter the sneutrino is the 
smaller the cross section becomes. The lightest stau mass 
$m_{\tilde\tau_1^{\pm}}$ is plotted in Fig.~\ref{fig:c8090sp}c. It is
the lightest of the charged sleptons and its mass decreases when 
$\tan\beta$ increases. This is due to the fact that stau mixing grows with
$\tan\beta$. Many of the curves are truncated because $\tilde\tau_1^{\pm}$
is too light. In Fig.~\ref{fig:c8090sp}d we have the mass of the lightest
up--type squark, which is mainly top--squark with a very small component of
charm--squark. Contrary to the previous case, the stop mass $m_{\tilde t_1}$ 
decreases when $\tan\beta$ decreases. This effect appears because the stop
mixing grows when $\tan\beta$ decreases and at the same time $\mu$ increases.

\begin{figure}
\centerline{\protect\hbox{\psfig{file=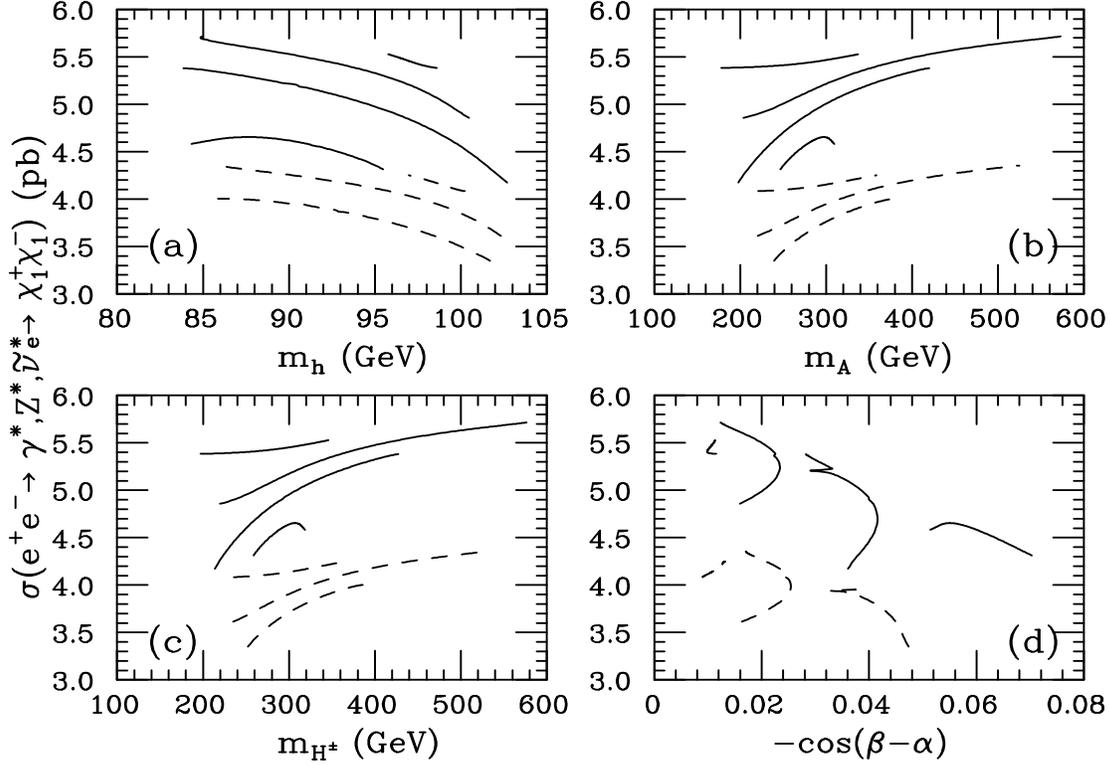,height=11cm,width=1.0\textwidth,angle=90}}}
\caption{
Relation between the total production cross section of two charginos 
$\tilde\chi_1^{\pm}$ and the following parameters of the Higgs sector: 
(a) lightest CP--even Higgs mass $m_h$, (b) CP--odd Higgs mass $m_A$, 
(c) charged Higgs mass $m_{H^{\pm}}$, and (d) the parameter 
$-\cos(\beta-\alpha)$.} 
\label{fig:c8090hs}
\end{figure} 
The Higgs sector is represented in Fig.~\ref{fig:c8090hs}. The lightest 
CP--even Higgs mass is plotted in Fig.~\ref{fig:c8090hs}a and satisfy 
$84<m_h<103$ and may be detected at LEP2 if the center of mass energy 
$\sqrt{s}=200$ GeV is achieved. In this case, if in addition to chargino
detection we have a measurement of the lightest Higgs mass, we can
determine the gluino mass with the aid of this figure. The CP--odd Higgs
is heavier than in the light selectron scenario of the previous section.
We plot $m_A$ in Fig.~\ref{fig:c8090hs}b, and it satisfy $170<m_A<580$
GeV. Strongly correlated with $m_A$ is the charged Higgs mass $m_{H^{\pm}}$
in Fig.~\ref{fig:c8090hs}c, which is slightly heavier: $200<m_{H^{\pm}}<580$
GeV. Consistent with the heaviness of the CP--odd Higgs, we find that the
Higgs sector is close to the decoupling limit, that is, the Higgs boson
$h$ behaves like the SM Higgs boson: $|\cos(\beta-\alpha)|<0.07$ as it can
be appreciated from Fig.~\ref{fig:c8090hs}d. 

\section{Conclusions}

In Supergravity models, with either the minimal SUGRA relation at the 
unification scale $A=B+m_0$ or the relation $B=2m_0$ motivated by 
solutions to the $\mu$--problem, we have shown that all the fundamental
parameters of the theory can be determined by the observation of
a pair of right--selectrons or a pair of light charginos. The necessary
experimental measurements are the mass of the observed particle, its 
total production cross section, and the mass of the lightest neutralino
which come from the decay of the observed particle. In the light selectron
scenario $m_0$ is small and $M_{1/2}$ may be large. On the contrary, in the
light chargino scenario, $M_{1/2}$ is small and $m_0$ may be large. In
this sense, the two scenarios complement each other. 

In the light selectron
scenario the neutral Higgs with $64<m_h<110$ GeV, and the lightest
stau with $45<m_{\tilde\tau_1^{\pm}}<85$ GeV, may be also produced at LEP2.
Therefore, a measurement of their masses can help in the determination of 
$\tan\beta$ due to the strong dependence of these masses on $\tan\beta$.
Analogously, in the light chargino scenario the neutral Higgs with
$84<m_h<103$ GeV, and the second lightest neutralino with 
$80<m_{\tilde\chi^0_2}<94$ GeV, may be also produced at LEP2. And a 
measurement of their masses can help in the determination of the gluino mass,
and with it the fundamental parameters of the theory.

In both scenarios, the allowed region in parameter space is rather small,
as it can be appreciated from Figs.~\ref{fig:s85ne} and~\ref{fig:c8090ne}.
Therefore, the detection of either right--selectrons or light charginos
will validate or ruled out the class of supergravity models analysed here.

\section*{Acknowledgements}

Part of this work was done in collaboration with Beatriz de Carlos and 
part with Steve F. King, to whom I express my appreciation. 
In Spain this work was supported by a DGICYT postdoctoral grant and by 
the TMR network grant ERBFMRXCT960090 of the European Union. Part of 
this work was done during my stay in the International Centre for 
Theoretical Physics, in Trieste, Italy.

\end{document}